\documentclass[a4paper,british,pra,twocolumn,superscriptaddress,nofootinbib,tightline]{revtex4-1}

\usepackage{xcolor}
\usepackage{hyperref}
\hypersetup{
    colorlinks,
    linkcolor={red!50!black},
    citecolor={blue!50!black},
    urlcolor={blue!80!black}
}

\usepackage[T1]{fontenc}
\usepackage[latin9]{inputenc}
\usepackage{color}
\usepackage{refstyle}
\usepackage{amsmath}
\usepackage{amssymb}
\usepackage{graphicx}
\usepackage{esint}

\graphicspath{{plots/}}

\makeatletter

\newcommand{\floor}[1]{\lfloor #1 \rfloor}
\newcommand{\ab}[1]{\left| #1 \right|}

\newcommand{\mean}[1]{\langle #1 \rangle}
\newcommand{\ket}[1]{| #1 \rangle}
\newcommand{\bra}[1]{\langle #1 |}

\newcommand{\beq}{\begin{eqnarray}}
\newcommand{\eeq}{\end{eqnarray}}

\newcommand{\eq}[1]{Eq.~(\ref{#1})}
\newcommand{\fig}[1]{Fig.~\ref{#1}}

\newcommand{\bs}[1]{\boldsymbol{#1}}

\newcommand{\proj}[1]{\ket{#1}\!\bra{#1}}

\definecolor{mygray}{gray}{0.5}
\definecolor{mymagenta}{cmyk}{0,1,0,0.12}

\AtBeginDocument{\providecommand\subref[1]{\ref{sub:#1}}}
\AtBeginDocument{\providecommand\figref[1]{\ref{fig:#1}}}
\pdfpageheight\paperheight
\pdfpagewidth\paperwidth

\RS@ifundefined{subref}
  {\def\RSsubtxt{section~}\newref{sub}{name = \RSsubtxt}}
  {}
\RS@ifundefined{thmref}
  {\def\RSthmtxt{theorem~}\newref{thm}{name = \RSthmtxt}}
  {}
\RS@ifundefined{lemref}
  {\def\RSlemtxt{lemma~}\newref{lem}{name = \RSlemtxt}}
  {}

\@ifundefined{textcolor}{}
{%
 \definecolor{BLACK}{gray}{0}
 \definecolor{WHITE}{gray}{1}
 \definecolor{RED}{rgb}{1,0,0}
 \definecolor{GREEN}{rgb}{0,1,0}
 \definecolor{BLUE}{rgb}{0,0,1}
 \definecolor{CYAN}{cmyk}{1,0,0,0}
 \definecolor{MAGENTA}{cmyk}{0,1,0,0}
 \definecolor{YELLOW}{cmyk}{0,0,1,0}
}

\usepackage{cancel}
\usepackage{braket}
\usepackage{tikz}

\makeatother

\usepackage{babel}

\begin{document}

\title{Proposal for a macroscopic test of local realism with phase-space measurements}

\author{Atul S. Arora}

\address{Indian Institute of Science Education \& Research (IISER), Mohali, Sector 81, Mohali 140 306, India}

\address{Naturwissenschaftlich-Technische Fakult\"at, Universit\"at Siegen, Walter-Flex-Str. 3, D-57068 Siegen, Germany}

\author{Ali Asadian}

\address{Naturwissenschaftlich-Technische Fakult\"at, Universit\"at Siegen, Walter-Flex-Str. 3, D-57068 Siegen, Germany}

\pacs{Draft}

\date{May-July 2015}
\begin{abstract}
We propose a test of local realism based on correlation measurements of continuum valued functions of positions and momenta, known as modular variables. 
The Wigner representations of these observables are bounded in phase space and therefore, the associated inequality holds for any state described by a non-negative Wigner function. This agrees with Bell's remark that positive Wigner functions, serving as a valid probability distribution over local (hidden) phase space coordinates, do not reveal non-locality. We construct a class of entangled states resulting in a violation of the inequality and thus truly demonstrate non-locality in phase space. The states can be realized through grating techniques in space-like separated interferometric setups. The non-locality is verified from the spatial correlation data that is collected from the screens.
\end{abstract}
\maketitle

\section{Introduction}

In 1935, Einstein, Podolsky, and Rosen (EPR) 
argued that the quantum-mechanical description of 
physical reality is not complete, and thus may be superseded by a more 
complete realistic theory which reproduces the quantum mechanical predictions, and 
at the same time, obeys the locality condition \cite{EPR}. Bell derived an experimentally testable inequality, in his seminal 1964 paper \cite{Bell}, which
bounds the correlations between bipartite measurements for any such local hidden-variable theory, but is violated
by quantum mechanics. This was a major breakthrough towards empirical 
tests of quantum mechanics against theories conforming to common sense. Since 
then, the results constraining the permissible types of hidden variable 
models of quantum mechanics have attracted much attention and have been 
reformulated as the problem of contextual measurements by Kochen and 
Specker~\cite{Kochen} and in terms of temporal correlations by Leggett 
and Garg~\cite{LG85}. Today, these concepts have mainly been formulated for intrinsic quantum degrees of freedom of microscopic particles such as spins, and tested in various 
experiments with photons~\cite{Aspect1999}, 
impurity spins~\cite{WaldherrPRL2011,KneeNatComm2012} or superconducting 
qubits~\cite{Palacios-LaloyNatPhys2010}. All experimental observations confirmed the validity of quantum mechanics at this level. 
 
The outstanding challenge is, however, to formulate similar tests with true phase space measurements, where non-locality is inferred directly from observing, the spatial degree of freedom, for example. 
This can be viewed as a natural extension to the macroscopic limit of local realism tests \cite{Reid}.  This is closer in spirit to the original EPR argument which uses phase space description, a natural concept in the classical world, to better address the reality and locality problems of quantum mechanics. 
Notably, the Wigner function associated with the entangled state used in the EPR argument, the so called EPR state, is non-negative everywhere \cite{Gneiting13}. That is why, Bell argued that EPR states do not lead to a violation of the inequalities derived from locality and hidden variable assumptions \cite{bellsBook}. This was because non-negative Wigner functions serve as valid joint probability distributions over local \emph{hidden} positions and momenta. Thereby, in principle, a model of local hidden variable can be attributed to such states. Banaszek and Wodkiewicz \cite{Parity98}, however, showed that using particular measurements, namely parity, EPR states can reveal non-local features indicating that not only the state itself but the type of correlation measurements is also important in any local realism test. This opens the discussion as to which measurements are good candidates for appropriately testing local hidden variable models in phase space \cite{Revzen, Son}. The problem of constructing a ``macroscopic'' test of local hidden variable models depends on choosing proper observables whose  Wigner representations satisfy the constraint imposed by the algebraic Clauser-Horne-Shimony-Holt (CHSH) \cite{CHSH69} expression. The term ``macroscopic'' henceforth, will be used to refer to the measurement of a particle's phase space coordinates. This class of measurements has a clear description in classical limit and its evaluation does not involve sharp measurements that betray ``quantum degrees of freedom''. 

The observables used here are the so called modular variables \cite{Ahar69}. Recently, such variables have found applications in detecting certain continuous variable (CV) entangled states ~\cite{Gneiting11, Barros, ModularExp} and quantum information \cite{ModularQI,Gris2014}. Furthermore, they have been used for fundamental tests such as macroscopic realism \cite{AsadianLG}, contextuality \cite{Plastino, AsadianPS} and even the GHZ test \cite{Massar01}. This strongly suggests that modular variables can be used for Bell inequality tests of local hidden variable theories as well. Recently a Bell test with discretized modular variables was proposed \cite{Ketterer15}. In the present work we put forward a Bell test with ``continuous'' modular variables which requires phase space measurements.

The paper is organized as follows.
In Section \ref{framework}, we introduce our framework for the Bell test, aiming to use most ``classical-like'' variables and measurements. This motivates a macroscopic test of local realism \cite{Reid}. In Section \ref{construction} we construct a Bell operator from modular variables, for which the violation is achieved only if the state is described by a negative Wigner function. We then proceed with identification of the relevant entangled state, explicitly showing the violation. 
Finally, in Sections \ref{measurementSchemes} and \ref{physicalImplementation} we show how the entire test can be implemented in a double (multi-slit) grating setup \cite{Gneiting13}. Grating techniques have been used to experimentally demonstrate quantum matter waves \cite{GratingExp}. We summarize by briefly discussing the outlook in Section \ref{discussion} and conclude in Section \ref{conclusion}. 

\section{Framework for a macroscopic Bell test \label{framework}}

In what follows we develop a test of local realism which complies with Bell's aforesaid argument. The central problem here is to construct a Bell-operator of the CHSH form
\begin{equation}
\label{Belloperator}
\hat {\mathcal{B}} \equiv \hat A_1\otimes (\hat A_2+\hat A_2') +\hat A_1'\otimes(\hat A_2-\hat A_2')
\end{equation}
expressed in terms of suitable local CV observables $\hat A_i$ which must be restricted to a limited range of values to impose a well-defined classical bound. 
We therefore require the observables to satisfy the following properties. 

(a) 
Eigenvalues of $\hat A_i$, $|a_i|\leq 1$, which for bounded observables can be achieved trivially by re-scaling. An example of this is the parity operator.\\
While this condition is enough to obtain a classical bound, we demand an extra constraint which is necessary for probing non-locality in phase space. 

(b) The observable $\hat A$ corresponds to a \emph{bounded} c-number function in phase space obtained from the Wigner-Weyl correspondence ($q\leftrightarrow \hat q$, $p\leftrightarrow \hat p$), viz.
\beq
|\mathcal{W}_{\hat A}(q, p)| \equiv \left| \int dq' e^{ipq'}\bra{q-\dfrac{q'}{2}}\hat A \ket{q+\dfrac{q'}{2}}\right| \leq 1.
\eeq 
This entails,
\begin{align*}
|\mathcal{W}_{\hat {\mathcal{B}}}(\bs q,\bs p)|&=|\mathcal{W}_{\hat A_1}(q_1,p_1)[\mathcal{W}_{\hat A_2}(q_2,p_2)+\mathcal{W}_{\hat A_2'}(q_2,p_2)] \\&+\mathcal{W}_{\hat A'_1}(q_1,p_1)[\mathcal{W}_{\hat A_2}(q_2,p_2)-\mathcal{W}_{\hat A_2'}(q_2,p_2)]|\leq 2,
\end{align*}
for the Wigner representation of the Bell operator where $\bs q\equiv(q_1,q_2)$ and  $\bs p\equiv(p_1, p_2)$.
Accordingly, for any state, including the EPR state, described by a valid (non-negative) probability distribution over phase space, the following inequality holds
\beq 
|\mean{\hat{\mathcal{B}}}|=\left| \int W_{\hat {\rho}}(\bs q,\bs p)\mathcal{W}_{\hat {\mathcal{B}}}(\bs q,\bs p)d\bs q d\bs p \right| \leq 2,
\eeq 
where $W_{\hat{\rho}}$ is the Wigner quasi-probability distribution corresponding to $\hat {\rho}$ given by $
W_{\hat {\rho}} = \mathcal{W}_{\hat {\rho}}/2\pi \hbar$. A violation therefore must necessarily arise from the negativity of the Wigner function describing the state.

Although formally valid, the Bell inequality expressed in terms of displaced parity operators used in Ref. \cite{Parity98,Lee09}, voids the second condition; their Wigner representations are given by delta functions which are unbounded in phase space. \\  
The measurement scheme used for evaluating the correlations, must have a clear classical limit for any reasonable ``macroscopic test''. This measurement strategy is in marked contrast with other approaches that use parity measurements \cite{Parity98}. Parity measurements unlike phase space measurements, require resolving intrinsic quantum degrees of freedom and thus have no classical analog.
It has been shown that for sufficiently sharp measurements the system inevitably enters a quantum regime and no classical description is possible \cite{Kofler07}. Thus such measurements remotely resemble ``classical-like'' measurements, if at all.

The binary binning of quadrature measurements has also been shown to be a possible scheme \cite{Wenger03,Monru99} where entangled Schr\"odinger Cat states (and their appropriate generalizations) are used. Here however, to preserve features characteristic of classical dynamics, we aim to adopt a different measurement strategy which retains the continuous spectra and uses phase space exclusively.

\subsection*{Phase space translation and modular variables }

One particular class of bounded observables can be constructed from the quantum mechanical space translation operator, $e^{-i\hat p L/\hbar}$. As its name suggests, this operator displaces a particle by a finite distance $L$, which in our case will be the distance between two adjacent slits. This operator is not an observable, therefore we define a symmetric combination   
\begin{align}
  \label{translation}
  \hat X \equiv\dfrac{e^{-i\hat p L/\hbar}+e^{i\hat p L/\hbar}}{2} =\cos(\hat p L/\hbar),
\end{align}
which is explicitly Hermitian and bounded by $\pm 1$. In fact the corresponding function $\ab{\mathcal W_{\hat X}}=\ab{\cos(pL/\hbar)}$ is also manifestly  $\le 1$. Further, when $\hat X$ is operated on $\ket{p}$, then only the modular part of $p$ is relevant to the value of the operator. Thus we may define 
\beq
\hat p_{\text{mod}\frac{h}{L}}\equiv(\hat pL/h-\floor{\hat pL/h})\dfrac{h}{L}
\eeq
and note that measuring $\hat p_{\text{mod}\frac{h}{L}}$ is sufficient for obtaining the value of $\hat X\equiv X(\hat p_{\text{mod}\frac{h}{L}})$. Conversely, measuring $\hat X$ only yields $\hat p_{\text{mod}\frac{h}{L}}$, not $\hat p$. 
The idea is to construct a Bell operator [see \eq{Belloperator}] from $\hat X$ in which the different measurement settings are chosen by transforming it using suitable unitary operators.

\section{The construction \label{construction}}

 Consider a localized state $\varphi(q) = \braket{q|\varphi}$ symmetric about the position $q=L/2$, where $L\equiv$length scale and $\varphi_n(q)\equiv \varphi(q-nL)$. 
We define 

\begin{equation*}
\left|\psi_{0}\right\rangle  \equiv  \frac{1}{\sqrt{M}}\sum_{n=-\floor{\frac{M}{2}}}^{\floor{\frac{M-1}{2}}}\left|\varphi_{2n+1}\right\rangle \  ,  \ 
\left|\psi_{1}\right\rangle  \equiv  \frac{1}{\sqrt{M}}\sum_{n=-\floor{\frac{M}{2}}}^{\floor{\frac{M-1}{2}}}\left|\varphi_{2n}\right\rangle.
\end{equation*}
\begin{figure}
\includegraphics[width=9cm]{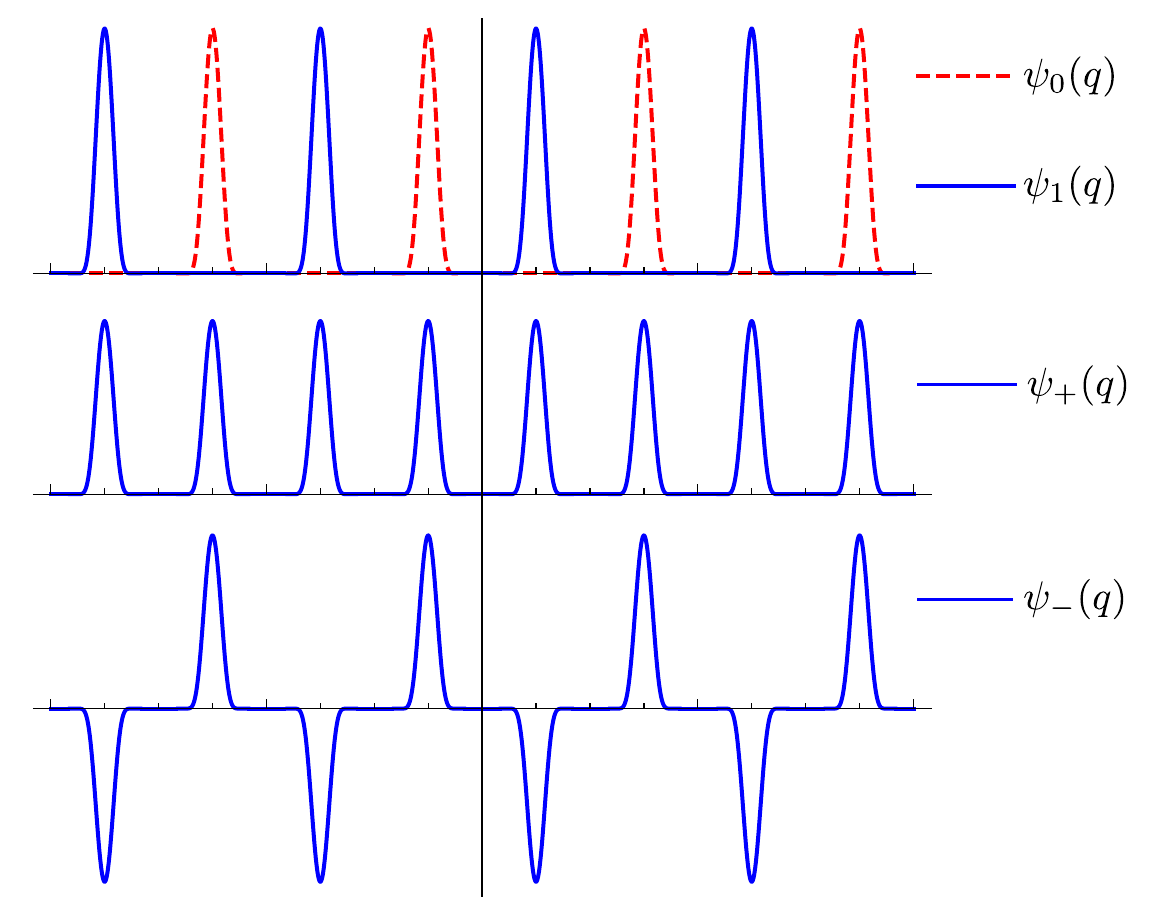}


\caption{(Color online) Illustration of multicomponent superposition states  $\left|\psi_{\pm}\right\rangle $ and $\left|\psi_{0}\right\rangle ,\,\left|\psi_{1}\right\rangle $
 for $N=8$. \label{fig:wavefns}}
\end{figure}
Using these states, as illustrated in \figref{wavefns}, we construct the states
\begin{equation}
\label{eqn:psipm}
\left|\psi_{+}\right\rangle \equiv\frac{\left|\psi_{0}\right\rangle +\left|\psi_{1}\right\rangle }{\sqrt{2}} \ \ , \ \
\left|\psi_{-}\right\rangle \equiv\frac{\left|\psi_{0}\right\rangle -\left|\psi_{1}\right\rangle }{\sqrt{2}}.
\end{equation}
These states were constructed with a partial translational symmetry which is appropriate to the bounded Hermitian operator $\hat X$ discussed earlier. These $N$-component superposition states can represent a 
delocalized particle after an $N$-slit grating. It follows that 
\begin{align*}
\bra{\psi_+} \hat X \ket{\psi_+} &= \frac{N-1}{N} \\
\bra{\psi_-} \hat X \ket{\psi_-} &= -\frac{N-1}{N},
\end{align*}
where $N \equiv 2M$ is the number of `slits'.
Before proceeding further, we introduce a unitary operator $\hat U$ to implement different measurement settings. 
Motivated by the spins we define $\hat U$ by its action 
\begin{align}
\label{eqn:U}
\hat{U}(\phi)\left|\psi_{0}\right\rangle   =  e^{i\phi/2}\left|\psi_{0}\right\rangle, \ \ \
\hat{U}(\phi)\left|\psi_{1}\right\rangle   = e^{-i\phi/2}\left|\psi_{1}\right\rangle.
\end{align}
More explicitly 
\[
\hat{U}(\phi)\equiv e^{i\hat{Z}\phi/2},
\]
where $\hat{Z}$ is s.t. $\hat{Z}\left|\psi_{0}\right\rangle =\left|\psi_{0}\right\rangle $ and $\hat{Z}\left|\psi_{1}\right\rangle =-\left|\psi_{1}\right\rangle $. We note that $\hat{Z}$ must differentiate between spatial wave-functions $\left\langle q|\varphi\right\rangle $ and $\left\langle q-L|\varphi\right\rangle $. It is thus natural to expect $\hat{Z}$ to be a function of  $\hat{q}_{\text{mod} 2L}$, i.e. 
$\hat{Z}\equiv Z(\hat{q}_{\text{mod}2L})$. For consistency then, we conclude that $Z$ must have the form of a square wave and define 
\[
\hat Z \equiv \rm sgn \left(\sin\dfrac{\hat q \pi}{L}\right).
\]

\begin{figure}
\includegraphics[width=8cm]{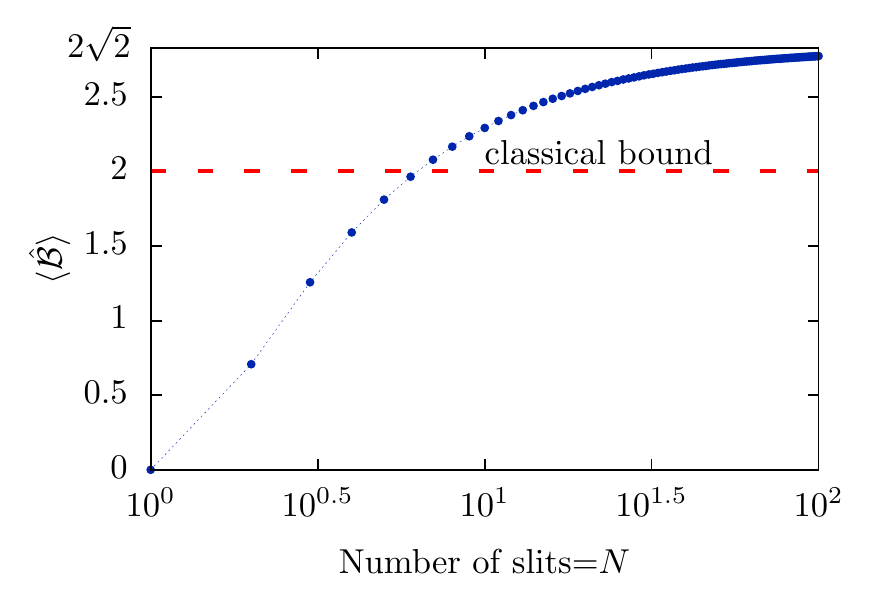}

\caption{(Color online) Practically, the number of slits, $N$ will be finite. The plot shows
$\braket{ \mathcal {\hat B} } $ as a function of $N$. To get a violation,
we need merely $8$ slits; with $50$ we almost saturate. \label{chshViolation}}
\end{figure}
The test is performed by considering two particles and their observers, Alice and Bob; they apply the aforesaid local unitaries to define the setting, and then measure $\hat X$. 
We claim that the suitable entangled state which will yield a violation, given this scheme, is
\begin{equation}
\label{entangledState}
\ket{\Psi} \equiv\frac{ \ket{\psi_{+}}_1\ket{\psi_{-}}_2 - \ket{\psi_{-}}_1\ket{\psi_{+}}_2 }{\sqrt{2}}.
\end{equation}
We now evaluate $\mean{\mathcal {\hat B}}$. This essentially requires terms like $ \mean{ \hat X(\phi) \otimes \hat X(\theta)}$, where $\hat X (\theta) \equiv \hat U^{\dagger}(\theta) \hat X \hat U(\theta)$. It can be shown that [see the Appendix] 
\beq
\mean{\hat X(\phi) \otimes \hat X(\theta)} = -\left( \frac{N-1}{N} \right)^2 \cos(\phi-\theta).
\eeq
Thus for particular angles, i.e. $\theta',\,\phi,\,\theta$ and $\phi'$ successively separated by $\pi/4$, we get
 \beq
\ab{\mean{ \mathcal {\hat B} }} =\left(\frac{N-1}{N}\right)^{2}2\sqrt{2}.
\eeq 
The violation, i.e.,  $\ab{\mean{ \mathcal {\hat B} }} >2$, requires $N > 6$; see \fig{chshViolation}. To interpret this, we must ensure that the assumptions of the framework are satisfied, viz. $\ab{\mathcal W_{\hat X(\phi)}}\le 1$. To that end, we note that
\beq
\ab{\mathcal W_{\hat X (\phi)}(q,p)} & = & \Big| \frac{1}{2}\int dq'e^{ipq'/\hbar} \bra{q-\frac{q'}{2}}\bigg( e^{-i\hat Z\phi/2} \nonumber \\
 &  & e^{i\hat{p}\frac{L}{\hbar}}e^{i\hat Z\phi/2} + \text{h.c.} \bigg) \ket{q+\frac{q'}{2}} \Big| \nonumber \\
 & = & \ab{\text{Re}\left( e^{-iZ_{-}(q)\phi/2}e^{ip\frac{L}{\hbar}}e^{iZ_{+}(q)\phi/2}\right)} \nonumber \\
 & = & \ab{\cos(pL/\hbar \pm Z_{\pm}(q)\phi)} \le 1,
\eeq
where $Z_{\pm}(q) \equiv Z[(q\pm\frac{L}{2})\text{mod}2L]$ and we used the fact that $Z(q)=-Z(q+L)$ (omitting the $\text{mod}2L$).

\subsection*{Passing Remarks} 

1. Pauli-like Commutation: While at first the definition of $\hat Z$ might appear arbitrary, we show that it naturally yields Pauli like algebra. We start with $[\hat{Z},e^{i\hat{p}L/\hbar}]$. 
To evaluate it, we multiply the second term with $\int dq\left|q\right\rangle \left\langle q\right|$ and obtain $\hat Ze^{i\hat{p}L/\hbar}+\hat Ze^{i\hat{p}L/\hbar}$ where we've used $Z(\hat{q}_{\text{mod}2L})=-Z(\left(\hat{q}\pm L\right)_{\text{mod}2L})$. 
\[
[\hat{Z},\hat{X}]=2\hat{Z}\hat{X}=-2i\hat{Y},
\]
where $\hat{Y} \equiv i\hat{Z}\hat{X}$. Here $i$ was introduced to ensure $\hat{Y}^{\dagger}=\hat{Y}$, since $\hat{X}^{\dagger}=\hat{X}$ and $\hat{Z}^{\dagger}=\hat{Z}$. Similarly  $\{\hat{Z},\hat{X}\}=0$. From the definition of $Y$ and the anti-commutation, $\{\hat{Y},\hat{X}\}=0$ and $\{\hat{Y},\hat{Z}\}=0$ also follow trivially. We may point out that while $\hat{Z}^{2}=1$, it is not a sum of a 2 state projector and $\hat{X}^{2}\neq1$ in general. This manifests in the following relations.
\begin{eqnarray*}
[\hat{X},\hat{Y}] & = & -2i\hat{Z}\hat{X}^{2}=-2i\hat{X}^{2}\hat{Z}\\{}
[\hat{Y},\hat{Z}] & = & -2i\hat{X}.
\end{eqnarray*}
It is apparent that the exact SU(2) algebra is not necessary to arrive at a violation.

2. Asymmetry in Z and X: Using an analogous momentum translation operator, the following can be derived from the definition of $\hat p$.
\[
e^{i\hat{p}u}e^{i\hat{q}v}=e^{i\hbar uv}e^{i\hat{q}v}e^{i\hat{p}u}.
\]
For appropriate choices of $u,v$, the translation operators can be made to commute or anti-commute. In the former case, it means that one can simultaneously measure modular position and momentum (which is in stark contrast with $\hat x$ and $\hat p$ measurements) and in the latter case, one can define Pauli matrix like commutation.
Considering the operator (non-Hermitian for simplicity) $\hat{X}=e^{i\hat{p}L/\hbar}$, defining $\hat{Z}=e^{i\hat{q}2\pi/2L}$ is more natural. They also follow the desired anti-commutation $\{\hat{X},\hat{Z}\}=0$ and we could define $\hat{Y}=\hat{iZ}\hat{X}$ to get a more natural generalization. The question is why did $\hat{Z}=Z(\hat{q}_{\text{mod}2L})$ appear in the analysis. The cause of this asymmetry hinges on the preferential treatment of position space. We could have constructed states of the form $\left|\psi_{0}\right\rangle =\sum_{n}\left|q+nd\right\rangle $ and used the natural definition of $\hat{Z}$ to obtain the violation. The issue is that this forces us to choose a countable superposition of position eigenkets as our desired state.\footnote{Such a state is strictly not even in the Hilbert space.} If we start with better defined and broader class of relevant states, $Z(\hat{q}_{\text{mod}2L})$ appears naturally.

3. Commutation and classical limit: It is well recognized and can be shown that there is a tight relation between non-locality and non-commutativity of operators. 
The violation occurs for choices of settings whose corresponding observables do not commute. In our construction we can demonstrate that the source of violation can be clearly attributed to the non-commutativity between position and momentum, $[\hat q,\hat p]=i\hbar$. This would be regarded as a further illustration that our approach provides a relevant test in phase space. We show that in our case $[\hat X(\theta),\hat X(\theta')]\neq 0$.
To prove that, we use $\hat X(\theta)=\hat{X}e^{i\hat{Z}\theta}$, $e^{i\hat{Z}\theta}=\cos\theta+i\hat{Z}\sin\theta$ and the previous results, to arrive at 
\begin{align*}
[\hat X(\theta),\hat X(\theta')] &=
2i\sin(\theta'-\theta) \hat Z \hat X^2 \\
&=2i \hat Z \hat X^2 \neq 0,
\end{align*}
where the last equality holds when the angles are as defined earlier. Classically this term not only vanishes, the different measurement settings also become identical. 
The Heisenberg equation of motion for the displacement operator 
\begin{align}
\frac{d\hat X}{dt} &=i\hbar^{-1} [\hat Z,\hat X] \nonumber \\  
  &=i\hbar^{-1} \Big( Z(\hat q_{\text{mod}2L})-Z(\hat q_{\text{mod}2L}\pm L) \Big) \hat X \nonumber \\
  &=i\hbar^{-1} 2 \hat Z \hat X
\end{align}
where $\hat Z$ is the potential, shows that $\hat X (\theta)$ is essentially $\hat X$ at some later time. However, classically, since the particle experiences no force (constant potential), $X(t)=X(t_0)$. This peculiarity is the same as that of the scalar Aharonov-Bohm effect, which is exploited here for realizing different measurement settings.
Manifestly then, the non-commutativity of $\hat q$ and $\hat p$ results in $\hat X(t)\neq \hat X(t_0)$ (as it follows a non-local equation of motion \cite{PopescuNatPhys2010}) which is pivotal for the violation.


\section{Measurement schemes \label{measurementSchemes}}

The scheme requires us to evaluate the correlation functions such as $\mean{\hat{X}(\theta)\otimes\hat{X}(\phi)}$. Equivalently, the measurement settings can be chosen by applying the corresponding local
unitaries  on the entangled state, that is $\ket{\Psi_{\theta\phi}}=\hat U(\theta)\otimes\hat U(\phi)\ket{\Psi}$. Therefore, obtaining $\left|\left\langle p{}_{1},p{}_{2}|\Psi_{\theta \phi}\right\rangle \right|^{2}$ is sufficient for evaluating 
 $\braket{ \hat{X}(\theta)\otimes\hat{X} (\phi)} = \int dp{}_{1}dp{}_{2}\cos(p{}_{1}L/\hbar)\cos(p{}_{2}L/\hbar)\left|\left\langle p{}_{1},p{}_{2}|\Psi_{\theta\phi}\right\rangle \right|^{2}$. 

It is known that in the far-field approximation \cite{Gneiting13}
\begin{equation}
\label{eqn:measure}
\left| \braket{p_1=\frac{p_zq_1}{D},p_2=\frac{p_zq_2}{D}|\Psi_{\theta\phi}} \right|^2=\dfrac{D^2}{p_z^2} \left| \braket{q_1,q_2 | \Psi_{\theta\phi}^\text{screen}} \right|^2,
\end{equation}
where $\left|\Psi{}_{\theta\phi}^\text{screen}\right\rangle $ is the state of the system at the screen, $D$ is the distance between the gratings
and the screens and $p_{z}$ is the $z$ component of momentum of the particle. For a photon, $p_{z}=h/\lambda$ while for a massive particle with mass $m$, $p_{z}=mD/T$, where $T$ is the time taken to arrive at the screen from the grating (see, Fig. \ref{fig:scheme}). The idea is simply that the momentum distribution at the grating can be recovered by observing the spatial distribution at the screen, sufficiently far away.

\section{Physical Implementation \label{physicalImplementation}}
\begin{figure*}
\includegraphics[width=17cm]{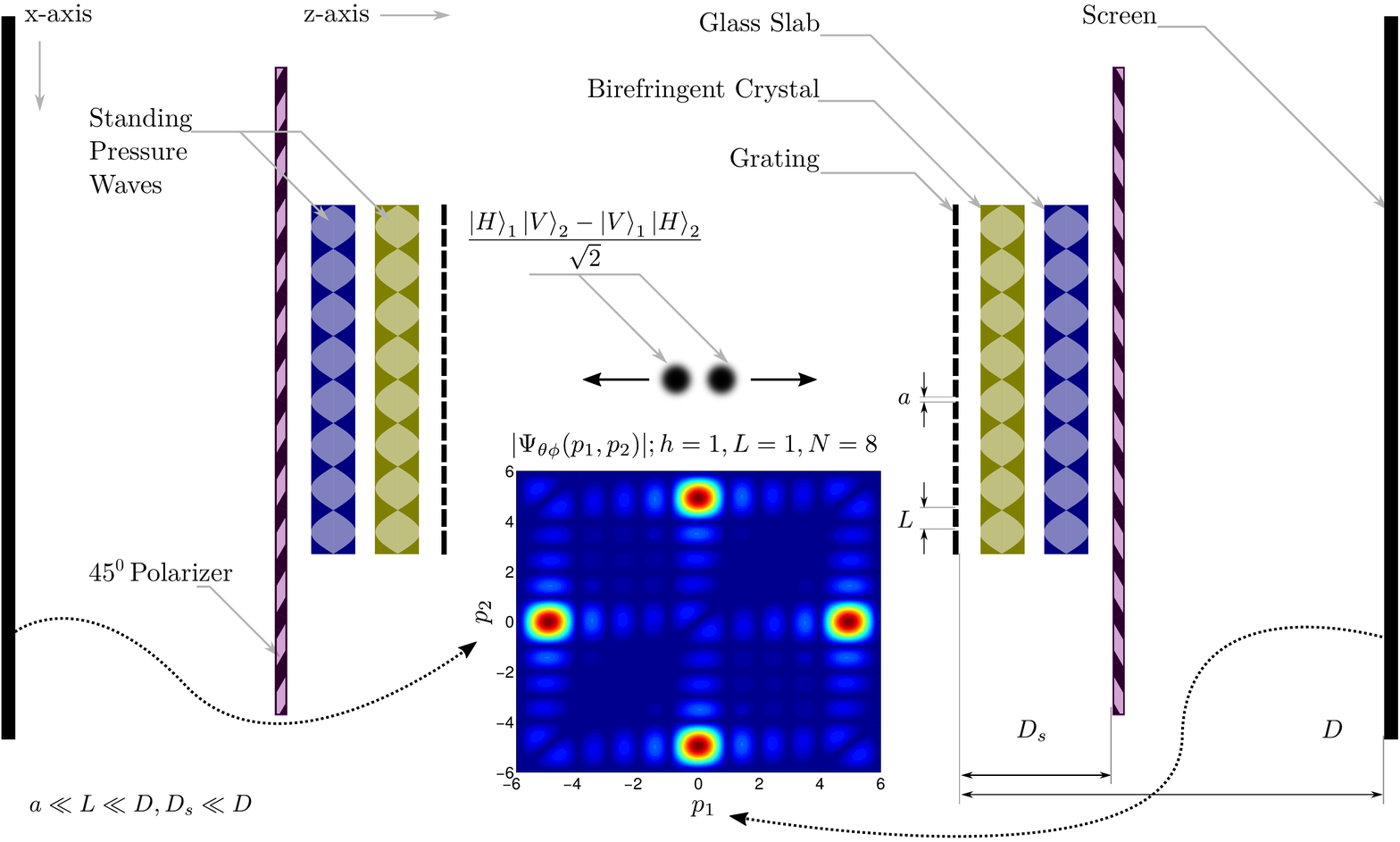}
\caption{(Color online) The experimental setup for implementing the test. It includes the scheme for creating the necessary entangled state. See the text for further details. \label{fig:scheme}}
\end{figure*}

The test can be implemented in a quantum interferometric setup, using grating techniques to create multi-component superposition states, as is done in matter wave experiments for instance. 
We show that this scheme can be implemented using photons. We harness the two degrees of freedom of a photon, it's polarization and it's spatial degree of freedom to construct the required state. With a slightly modified setup, it is possible to do the same with spin and position
for matter waves (see \subref{c7}). The final setup is given in \figref{scheme}. We need only consider the quantum mechanical description along the $x$-axis.

\subsection{Creation of the entangled state}

The desired entangled state is $\ket{\Psi}$, as stated in \eq{entangledState}. 
We start with noting the triviality of constructing a $\ket{\psi_{+}}$
 state (see \eq{eqn:psipm}). Consider a source that produces a state $\ket{\gamma}$ at the grating. $\braket{q | \gamma}$ is assumed to be a real Gaussian with $\sigma\gg 2NL$. The grating has $N$ slits of width $a\ll L$, separated by a distance $L$ (center to center). After the grating, we obtain $\ket{\psi_+}=\hat G\ket{\gamma}$, where $\hat G$ maybe formally defined accordingly. Similarly the $\ket{\psi_{-}}$
 state can be constructed by using glass slabs at alternate slits, such that the phase introduced is $\pi$. 
In \figref{scheme}, if you consider only one particle, and disregard
everything after the grating, then the setup is expected to produce a $\left|\psi_{+}\right\rangle $ state, right after it. To produce the desired entangled state, we start with two entangled photons, such that their polarization state can be expressed as $\ket{\chi} \equiv \frac{ \ket{H}_1\ket{V}_2 -\ket{V}_1\ket{H}_2 }{\sqrt{2}}$. Their spatial description (along $x$-axis) is initially assumed to be $\ket{\gamma}_1 \ket{\gamma}_2$ so that the post grating state is

\[
\frac{ \ket{H}_1\ket{V}_2 -\ket{V}_1\ket{H}_2 }{\sqrt{2}} \ket{\psi_+}_1\ket{\psi_+}_2.
\]
If we had glass slabs, whose refractive index (given some orientation) was say $\eta_{H}=1$ for a horizontally polarized beam and $\eta_{V}=\eta\neq1$ for vertical polarization
, then we could harness the entangled polarization state to create the required spatially entangled state. Birefringent crystals have such polarization dependent refractive indices. Assume that alternating birefringent crystals have been placed after both the gratings with appropriate thickness so that 
the subsequent state is

\[
\frac{\ket{H}_1\ket{V}_2\ket{\psi_+}_1\ket{\psi_-}_2 - \ket{V}_1\ket{H}_2\ket{\psi_-}_1\ket{\psi_+}_2}{\sqrt 2}.
\]
If the polarization state is traced out, the resultant state will be mixed, hence useless. 
Instead, a $45^{\circ}$ polarizer is introduced after which (see \subref{c8}) the target entangled state
\[
\ket{\chi_{45}}\ket{\Psi} = \ket{\nearrow}_1\ket{\nearrow}_2 \frac{\ket{\psi_+}_1\ket{\psi_-}_2 - \ket{\psi_-}_1\ket{\psi_+}_2}{\sqrt 2}
\]
is obtained, where $\left|\nearrow\right\rangle \equiv\left(\left|H\right\rangle +\left|V\right\rangle \right)/\sqrt{2}$. As a remark, it maybe be stated that although to arrive at this result we assumed that $\eta_{H}=1$, which is unreasonable physically, we can compensate for $\eta_{H}\neq1$ by putting appropriate glass slabs at the alternate empty slits, to produce zero relative phase when the polarization is horizontal.

\subsection{Measurement Settings}

The measurement setting is applied by local unitaries like $\hat U(\theta)\otimes \hat U(\phi)$. A local unitary can be performed by placing alternating glass slabs of widths such that \eq{eqn:U} holds. These slabs may be placed right after the birefringent crystals, before the polarizer. 
The final state just after the polarizer is given by $\left|\Psi_{\theta\phi}\right\rangle=\hat U(\theta)\otimes \hat U(\phi)\left|\Psi\right\rangle $, where $\theta\phi$ is one of the four possible measurement settings. 


\subsection{Effective Practical Setup}

Placing glass slabs may not be suitable for fine gratings, although a similar setup maybe possible \cite{Farias15}. Practically we can implement the same scheme using the setup shown in \figref{scheme}. 
The first large slab is a Birefringent crystal ($\eta_H,\eta_V$) while the adjacent slab is plain glass ($\eta$). We generate longitudinal standing pressure waves so that the effective thickness at alternate grating sites are given by $d_0,d_1$ and $l_0,l_1$ for the crystal and slab respectively. The phase difference between a horizontal $\ket{\psi_0}$ and $\ket{\psi_1}$ will be given by $\eta_H(d_0 - d_1)\equiv \phi_A$; note that physically only phase differences are essential. For the vertical component, it'll be $\eta_V(d_0-d_1)$. If we impose $\eta_V(d_0-d_1)=\pi+\phi_A$, then we would've created\footnote{up to an overall phase} the state
 \[ e^{i\hat Z \frac{\phi_A}{2}}\frac{\ket{\psi_+} + \ket{\psi_-}}{\sqrt{2}} \]
for an incident $\frac{\ket{H}+\ket{V}}{\sqrt{2}}$ polarization state. $d_0$ and $d_1$ will be constrained by some relation depending on physical properties of the crystal; they'll also depend on the amplitude of the longitudinal wave. From this and the imposed constrain, $d_0,d_1$ and the corresponding amplitude can be determined. However, we have not the freedom to change $\phi_A$. To remedy this, we use the glass slab. It will introduce an additional relative phase $\eta(l_0-l_1)\equiv \phi_B$. Here, again $l_0,l_1$, may satisfy some constraint, but will depend on the amplitude which is adjustable. Thus, by changing this amplitude, we can set the relative phase $\phi=\phi_A+\phi_B$ arbitrarily.

Spatial light modulators maybe used to more conveniently implement the aforesaid action of the glass slab and Birefringent crystal.

Effectively therefore, this scheme allows for both creation of the entangled state and changing the measurement settings in a practical way.

\section{Discussion \label{discussion}}

It is worth adding that one can use an alternative measurement strategy giving the same violation of the inequality. That is, to measure the modular variable with two-valued POVM elements $\hat E_\pm$, given by
\begin{equation}
\hat E_\pm=\dfrac{1}{2}  (\hat{\mathbb{I}}\pm\hat X),
\end{equation}
satisfying $\hat E_++ \hat E_-=\hat {\mathbb{I}}$. It follows that $\mean{\hat X}=p_+-p_-$ where the probabilities of getting $\pm$ outcomes, $p_\pm=\mean{\hat E_\pm}$, can be determined from the observed binary statistics read out from an ancillary two-level system \cite{Horodecki}. 

An interesting problem is to develop our approach to finite-dimensional systems, qudits. A class of Bell inequalities was proposed by Collins \emph{et al.} \cite{Collins}, which is useful for demonstrating nonlocality in high-dimensional entangled states.  For our version of Bell inequality generalized to $d$-dimensional systems can be achieved by using the discrete translation operators known as Heisenberg-Weyl  or Generalized Pauli operators, i.e., $e^{-i2\pi \hat P l/d}$, whose action is $e^{-i2\pi \hat P l/d}\ket{n}=\ket{n+l}$ where $l$ describes the steps translated in discrete position space with periodic boundary conditions and  $\hat P=\sum_{k=0}^{d-1} k\proj{k}$ is the discrete momentum operator. From this we obtain the relevant discrete modular variable $\hat X^l_d=\cos (2\pi \hat P l/d)$. 
We expect that the class of $d$-dimensional entangled states which demonstrate nonlocality here will be different from those considered by Collins \emph{et al.} \cite{Collins} and Lee \emph{et al.} \cite{LeeHigh09}.

It is obvious from the properties of the modular variables we use, that the violation is more pronounced for higher number of slits. One can however imagine that those entangled states created with slits fewer than the minimum number needed for obtaining a violation, must also hold non-local properties. To reveal the non-locality in this range one may need a more optimal set of observables, which involve a suitable combination of different modular variables, as opposed to the set considered here. 

\section{Conclusion \label{conclusion}}
In the present work, we constructed a new Bell-operator in terms of phase space measurements via modular variables. In this scheme there is no possibility for bipartite system with  positive definite Wigner function, formally entangled or not, to yield a violation of the inequality. Therefore, a violation of the inequality truly contradicts local (hidden) phase space models. From this perspective, our scheme is strongly different from the other approaches reported in Refs. \cite{Parity98, Chen02,ParityPS,Vitell} where sharp quantum measurements with no classical analog have been used. The measurement observables in our scheme instead are very simple with a clear classical limit. The relevant entangled states used for achieving a violation of the inequality however required creation of multi-component superposition states characterized by negative Wigner function. Interestingly our scheme also involves the scalar Aharonov-Bohm effect, manifesting another type of nonlocality \cite{PopescuNatPhys2010}.


\section{Acknowledgements}
The authors acknowledge discussions with O. G\"uhne, C. Budroni and P. Rabl. A. S. A is grateful to O. G\"uhne and his group for their support and hospitality during his visit to the Universit\"at Siegen. The authors acknowledge the financial support from the German Academic Exchange Service (DAAD), the FQXi Fund (Silicon Valley Community Foundation) and also the KVPY programme, Department of Science and Technology, India; The DFG and the Austrian Science Fund (FWF) through Erwin Schr\"odinger Stipendium No. J3653-N27 are also acknowledged.

\section{Appendix}



\subsection*{Claims}

Here, we provide more detailed derivations of the results.

\subsubsection{Useful Expectation Values\label{sub:c4}}

$\left\langle \psi_{+}\left|\hat{X}\right|\psi_{+}\right\rangle =\frac{N-1}{N}$,
$\left\langle \psi_{-}\left|\hat{X}\right|\psi_{-}\right\rangle =-\frac{N-1}{N}$\\
$\left\langle \psi_{0}\left|\hat{X}\right|\psi_{0}\right\rangle =0$,
$\left\langle \psi_{1}\left|\hat{X}\right|\psi_{1}\right\rangle =0$\\
$\left\langle \psi_{1}\left|\hat{X}\right|\psi_{0}\right\rangle =\frac{\frac{N-1}{N}+\frac{N}{N}}{2}=\frac{2N-1}{2N}=\left\langle \psi_{0}\left|\hat{X}\right|\psi_{1}\right\rangle $\\
$\left\langle \psi_{-}\left|\hat{X}\right|\psi_{+}\right\rangle =\frac{-\left\langle \psi_{1}\left|\hat{X}\right|\psi_{0}\right\rangle +\left\langle \psi_{0}\left|\hat{X}\right|\psi_{1}\right\rangle }{2}=0=\left\langle \psi_{+}\left|\hat{X}\right|\psi_{-}\right\rangle $\\
\begin{eqnarray*}
\left\langle \Psi\left|\hat{X}\otimes\hat{X}\right|\Psi\right\rangle  & = & \frac{1}{2}\bigg(\left\langle \psi_{-}\left|\hat{X}\right|\psi_{-}\right\rangle \left\langle \psi_{+}\left|\hat{X}\right|\psi_{+}\right\rangle +\\
 &  & \left\langle \psi_{+}\left|\hat{X}\right|\psi_{+}\right\rangle \left\langle \psi_{-}\left|\hat{X}\right|\psi_{-}\right\rangle \bigg)\\
 & = & -\left(\frac{N-1}{N}\right)^{2}
\end{eqnarray*}

\subsubsection{For Arbitrary $\theta_{i}$ and $\phi_{i}$ $\left\langle \hat{U}^{\dagger}(\phi_{i})\hat{X}\hat{U}(\phi_{i})\otimes\hat{U}^{\dagger}(\theta_{i})\hat{X}\hat{U}(\theta_{i})\right\rangle =-\left(\frac{N-1}{N}\right)^{2}\cos(\phi_{i}-\theta_{i})$\label{sub:c5}}

Proof: We start with defining $\phi\equiv\phi_{i}$ , $\theta\equiv\theta_{i}$,
$\delta\equiv\phi-\theta$, $\delta'\equiv\delta/2$. Next, we note
that $\text{LHS}=\left\langle \Psi'\left|\hat{X}\otimes\hat{X}\right|\Psi'\right\rangle $
where $\left|\Psi'\right\rangle =\hat{U}(\phi_{i})\otimes\hat{U}(\theta_{i})\left|\Psi\right\rangle $.
\begin{eqnarray*}
\left|\Psi'\right\rangle  & = & \frac{e^{i\delta'}}{\sqrt{2}}\left(\frac{\left|\psi_{+}\right\rangle -\left|\psi_{-}\right\rangle }{\sqrt{2}}\right)\left(\frac{\left|\psi_{+}\right\rangle +\left|\psi_{-}\right\rangle }{\sqrt{2}}\right)\\
 &  & -\frac{e^{-i\delta'}}{\sqrt{2}}\left(\frac{\left|\psi_{+}\right\rangle +\left|\psi_{-}\right\rangle }{\sqrt{2}}\right)\left(\frac{\left|\psi_{+}\right\rangle -\left|\psi_{-}\right\rangle }{\sqrt{2}}\right)\\
 & = & \frac{e^{i\delta'}}{2\sqrt{2}}\left(\left|\psi_{+}\psi_{+}\right\rangle +\left|\psi_{+}\psi_{-}\right\rangle -\left|\psi_{-}\psi_{+}\right\rangle -\left|\psi_{-}\psi_{-}\right\rangle \right)\\
 &  & -\frac{e^{-i\delta'}}{2\sqrt{2}}\left(\left|\psi_{+}\psi_{+}\right\rangle -\left|\psi_{+}\psi_{-}\right\rangle +\left|\psi_{-}\psi_{+}\right\rangle -\left|\psi_{-}\psi_{-}\right\rangle \right)\\
 & = & \frac{e^{i\delta'}-e^{-i\delta'}}{2\sqrt{2}}\left|\psi_{+}\psi_{+}\right\rangle +\frac{e^{i\delta'}+e^{-i\delta'}}{2\sqrt{2}}\left|\psi_{+}\psi_{-}\right\rangle \\
 &  & -\left(\frac{e^{i\delta'}+e^{-i\delta'}}{2\sqrt{2}}\right)\left|\psi_{-}\psi_{+}\right\rangle -\left(\frac{e^{i\delta'}-e^{-i\delta'}}{2\sqrt{2}}\right)\left|\psi_{-}\psi_{-}\right\rangle 
\end{eqnarray*}
Now using \subref{c4}, we have
\begin{eqnarray*}
\text{LHS} & = & \left\langle \Psi'\left|\hat{X}\otimes\hat{X}\right|\Psi'\right\rangle \\
 & = & \frac{1}{2}\left(\frac{N-1}{N}\right)^{2}\Bigg[\left|\frac{e^{i\delta'}-e^{-i\delta'}}{2}\right|^{2}\\
 &  & -\left|\frac{e^{i\delta'}+e^{-i\delta'}}{2}\right|^{2}-\left|\frac{e^{i\delta'}+e^{-i\delta'}}{2}\right|^{2}\\
 &  & +\left|\frac{e^{i\delta'}-e^{-i\delta'}}{2}\right|^{2}\Bigg]\\
 & = & -\left(\frac{N-1}{N}\right)^{2}\frac{1}{2}\left[2\left(\cos^{2}\delta/2-\sin^{2}\delta/2\right)\right]\\
 & = & -\left(\frac{N-1}{N}\right)^{2}\cos\left(\delta\right)
\end{eqnarray*}

\subsubsection{Physical implementation with electrons is also possible\label{sub:c7}}

If we can show that the basic components used to describe the photon
setup can be translated to the electron setup, then in principle we
are through. (a) Glass slab: The equivalent is the electric AB effect.
We need to simply put a capacitor after the slit and the two components
will pick up a phase difference. (b) Polarizer: The Stern Gerlach
setup is the classic analogue. We simply block the orthogonal component.
(c) Birefringent crystal: This is slightly tricky. It can be modeled
by using a combination of gradient of magnetic field (as in Stern
Gerlach) and a capacitor. We start with an equivalent superposition
of spin states, $\frac{\left|\uparrow\downarrow\right\rangle -\left|\downarrow\uparrow\right\rangle }{\sqrt{2}}\left|\psi_{+}\psi_{+}\right\rangle $.
To construct the spin dependent $\left|\psi_{-}\right\rangle $ state,
we use the magnetic field gradient to spatially separate the $\left|\uparrow\right\rangle $
and $\left|\downarrow\right\rangle $ states. We place capacitors
as described at the spatial position corresponding to $\left|\downarrow\right\rangle $
say. Thereafter, we remove the magnetic field gradient and allow the
beams to meet again. This will effectively act as a Birefringent crystal,
since the phase difference is spin dependent.

\subsubsection{Action of a polarizer\label{sub:c8}}

If we define $\left|\nearrow\right\rangle \equiv\frac{\left|H\right\rangle +\left|V\right\rangle }{\sqrt{2}},$
$\left|\nwarrow\right\rangle \equiv-\frac{\left|H\right\rangle -\left|V\right\rangle }{\sqrt{2}}$
and the $45^{\circ}$ projector as $\left|\nearrow\right\rangle \left\langle \nearrow\right|$,
then both $\left|H\right\rangle \to\left|\nearrow\right\rangle $ and $\left|V\right\rangle \to\left|\nearrow\right\rangle$
where of course with a probability $1/2$, the photon will be lost.

\subsubsection{More on measurement\label{sub:c9}}


It is essential to know what ballpark resolution is required for detecting the violation from the screen. We note
\[
\ab{\braket{p_1, p_2 | \Psi_{\theta \phi}}}=\ab{\tilde{\varphi}(p_1) \tilde{\varphi}(p_2) F_{\theta \phi}(p_1,p_2)}
\]
where
\begin{align*}
F_{\theta \phi}(p_1,p_2)&=\frac{1}{\sqrt 2} \sum_{n,m = -\floor{\frac{M}{2}} }^{\floor{\frac{M-1}{2}}} e^{i(np_1 + mp_2)L/\hbar} \\
& \Big[ -\cos(\delta') [(-1)^m - (-1)^n] \\
+ &i \sin(\delta') [1 + (-1)^{n+m}] \Big]
\end{align*}

$\tilde{\varphi}(p) \equiv \braket{p|\varphi}$ and $\delta' = (\phi - \theta)/2$. 
Since the wave-function $\varphi(q)$ was assumed sharp with respect to $L$, $\ab{ \tilde \varphi(p)}$ will only correspond to a broad envelope, over the range $(-Nh/2L,Nh/2L)$. Thus the main feature of $\ab{\braket{p_1,p_2,\Psi_{\theta \phi}}}^2$ will be given by $\ab{F_{\theta \phi}}$ as shown in \fig{fig:scheme}. Graphically it is clear that resolving at the scale $p_{\text{typ}}=\frac{h}{L}$ should be sufficient to capture the relevant features. On the screen, this translates to a typical length, $q_{\text{typ}}=\lambda D /L$ which follows from \eq{eqn:measure} and $p_z=h/\lambda$ for a photon. This is reminiscent of typical diffraction experiments and is in units that are readily measurable.


\bibliography{Modular}

\end{document}